**Deterministic Chaos in Radioactive Decay**

Július Krempaský[(a)], Štefan Húšťava[(b)], Pavol Valko[(a)*]

[(a)]*Department of Physics, Faculty of Electrical Engineering and Information Technology, Slovak Technical University, Ilkovičova 3, 812 19 Bratislava, Slovak Republic*

[(b)]*Department of Physics, Faculty of Civil Engineering, Slovak Technical University, Radlinského 11, 813 68 Bratislava, Slovak Republic*

**Abstract**

We examine the radioactive decay of iodine in terms of its Kolmogorov entropy, observing a consistency with the presence of a regime of deterministic chaos in the vacuum dynamics.

**1. Introduction**

The direct observation that the universe appears to be expanding at an increasing rate [1], and Wilkinson Anisotropy Probe observation [2], fit well with the assumption that the universe is dominated by dark energy in the form of a cosmological constant. Although the nature of dark energy is currently unknown, it is often associated with the properties of the physical vacuum. Among the various models of a possible physical vacuum structure and its associated energy, a fractal-like structure has been proposed [3,4].

The expression "fractal structure" can be understood within two different contexts. The first is the real world, in which objects such as trees, blood circulatory system, lungs, etc. demonstrate a fractal structure. Their characteristic feature is a so-called self-similarity as described in Ref. [5]. The second is in a phase space, wherein the structure is a result of dynamics which is able to modify a simple attractor system into a "strange attractor" system accompanied by a regime of deterministic chaos [6,7]. The existence of such non-random chaotic dynamics in climate systems has been shown by Lorenz [8] and others [9,10]. A parameter providing a measure of a deterministic chaos presence in a dynamical system is the maximal Lyapunov exponent ($\lambda_{max}$), for which $\lambda_{max} > 0$ is found for systems exhibiting a deterministically chaotic behaviour and $\lambda_{max} = 0$ for "usual" stochastic chaos caused by a large number of degrees of freedom. Further details concerning deterministic noise and characteristic exponents in dynamical systems are given in Ref. [11].

Although the influence of a fractal-like vacuum structure should be imprinted on all physical processes, it is natural to expect that the observable effects are many orders of magnitude below our experimental possibilities. Typical systems where one could possibly search for such signatures are those with extreme mass-energy concentrations, like compact object (neutron star or black hole) mergers. Smaller but still very high mass-energy density objects are atomic nuclei.

It is evident that direct observation of a physical vacuum structure is not possible at this time, neither highly probable in the future. The sufficient requisite of nontrivial vacuum dynamics would be to determine, for some process influenced by the vacuum, whether it belongs to the category of deterministic chaos or to simple stochastic chaos. The decay of radioactive nuclei has previously been experimentally examined for deviations from exponentiality and possible "new-physics" [12,13]. We here examine the Kolmogorov entropy of radioactive decay as an indicator of a possible fractal-like dynamics of the underlying physical vacuum, using a time series of more than $3 \times 10^5$ individual decay measurements of radioactive $^{131}$I.

## 2. Experimental and basic data treatment

A radioactive source of $^{131}$I in liquid form as a solution of sodium iodide in water was used. Only gamma radiation from the daughter $^{131}$Xe was measured, using a proportional counter (NPGD 02, Bitt Technology) in an energy window of 60-1600 keV. The bias (amplifier) stability was 50 ppm, with a bias voltage source stability of 6‰ at 1700 V. The count-rate stability was better than 1%, with an overall detection efficiency of 1%. The measurement duration for each data segment was 1 s, with the observed counting rate continuously decreasing from approx. 1600 to 1100 events/s due to the relatively short $^{131}$I lifetime. The average background in the selected energy window was 4 events/sec.

The raw experimental signal is not appropriate for the intended cross correlation search between the individual measurements, due to the decreasing number of events in the intervals. Application of a correlation search method to such a signal would generate artificial correlation sums in the form of step-like functions. The data was therefore first transformed into a quasi-stationary dataset by removing the time-dependent part of the signal via fitting the measurements with an exponential decay. The fitted half-life was $T_{1/2} = (710.62 \pm 0.39) \times 10^3$ s, i.e. slightly longer than the tabulated value [14] for the $^{131}$I β-decay source, but in excellent agreement with the source populating several $^{131}$Xe excited levels, including the longer-lived (11.84 day) 163.9 keV level.

The differences between experimental event rates and the fit value, or residues, showed maximal excesses of up to 174 events, and deficits of up to -166 events, over totals of > 1000 events.

If the data treatment modified the dataset properties, the residue average should deviate from zero: while the average is $-4.7 \times 10^{-4}$ with a most probable deviation of $6.8 \times 10^{-2}$, i.e. the data treatment up to this point satisfies the requirement of preserving the statistical properties of the original dataset.

## 3. Correlations in analyzed signal and deduced Kolmogorov entropy

Several methods to distinguish between the two kinds of chaos exist. They are based on either a direct evaluation of $\lambda_{max}$ [15], or on attractor reconstruction techniques, mainly implemented in Belousov-Zhabotinskii chemical reactions dynamics studies [16,17]. Here we follow an alternative method based on correlated statistical measure. The correlation integral C(r) is defined as [18,19]

$$C(r) = \lim_{N \to \infty} \frac{1}{N^2} \sum_{i,j=1}^{N} \Theta(r - |\vec{X}_i - \vec{X}_j|), \qquad (1)$$

where $\Theta(r)$ is a Heaviside function. The correlation integral $C(r)$ is also proportional to the power of r as

$$C(r) \sim r^\nu. \qquad (2)$$

where ν is closely related to the fractal (Hausdorf) dimensionality $D$ as $\nu \leq D$.

For real physical processes, further refinement of the correlation integral, based on Kolmogorov entropy ($K$) estimates, has been proposed [20]. This approach distinguishes random ($K = \infty$), deterministically chaotic (constant $K \neq 0$), and ordered ($K = 0$) systems directly from experimental data represented by a time series of individual measurements, assuming

$$C_d(r) \underset{\substack{r \to 0 \\ d \to \infty}}{\sim} r^\nu \exp(-d\tau K_2) , \qquad (3)$$

where the parameter $K_2 < K$, $d$ is an integer correlation order, and $\tau$ is the time separation between individual data values.

Correlation integrals $C_d(\varepsilon)$ for the refined dataset were calculated as:

$$C_d(\varepsilon) = \frac{1}{N^2} \left\{ \text{number of pairs } (n,m) \text{ with } \sqrt{(x_n - x_m)^2 + (x_{n+1} - x_{m+1})^2 + \ldots + (x_{n+d-1} - x_{m+d-1})^2} < \varepsilon \right\} \qquad (4)$$

where $x_i$ represent the number of events in each interval, and $\varepsilon = 1,2,\ldots x_{max}$. To simplify numerical part of analysis the whole studied dataset was shifted towards positive integer values with $x_{max}=340$ in our case.

The results of the individual correlation integrals calculated for the dataset, presented as a function of $\varepsilon$, are shown in Fig. 1. We obtain the expected random correlation at low $\varepsilon$, but also a series of straight lines in the scaling region ($\varepsilon > 240$) with a slope of $\nu = 2.19 \pm 0.20$. This represents a lower limit for the fractal dimensionality of the system [18].

The $K$ is estimated following Ref. [20]: the set of 2$^{nd}$ order Renyi entropies $K_{2,d}$ was computed as

$$K_{2,d}(\varepsilon) = \frac{1}{2} \ln \frac{C_d(\varepsilon)}{C_{d+2}(\varepsilon)} , \qquad (5)$$

which are related to $K_2$ as

$$\lim_{\substack{d \to \infty \\ \varepsilon \to 0}} K_{2,d}(\varepsilon) \sim K_2 . \qquad (6)$$

To approximate $K_2$ according to Eq. (6), we use the Renyi entropy calculated at $\ln(\varepsilon) = 5.5$, i.e. at the lowest point of the scaling region. As apparent from Fig. 2, $K_{2,d}$ at this point converges towards $K_2 \sim 0.1$. Since both the correlation sums and consequently also the calculated $K_{2,d}$ values are exact numbers, the possible error in $K_2$ is estimated from the spread of values at the highest dimension $d$ to be ~ 0.03. Observed $K_2 > 0$ represents a sufficient condition for the presence of a deterministic chaotic mode in the system [20], which is the central point of the analysis.

## 4. Discussion

Our calculated parameters $\nu$ is also surprisingly close to the corresponding parameter of a Mackey-Glass system, described by a nonlinear time delay ($\tau = 23$) differential equation ($\nu = 2.4$) [20] and known to possess complex dynamics including a regime of deterministic chaos. It is also near the above-mentioned Lorenz climate system [8] with the Ljapunov exponents ($\lambda_i$) derived from the Kaplan-Yorke equation [21] $D = 2.05 \pm 0.01$ (for 3 dimensional systems).

Our analysis of a radioactive decay process, based on an event number correlation search between individual measurements, supports the idea that the process is not entirely random (uncorrelated), and in fact exhibits the presence of a deterministic chaotic mode. It is important to stress, however, that the calculated $\nu$ and $K_2$ might be severely influenced by many sources of error not associated with radioactive decay itself. Any finite nonzero $K_2$ value could be produced as a result of the combined effect of purely deterministic ($K_2 = 0$) and purely random ($K_2 = \infty$) processes. Unavoidable variations of the detector amplifier gain (or bias voltage), in a predictable repeatable pattern, would artificially increase the calculated correlation. Conversely, the influence of any purely stochastic noise source (e.g. a purely random background) would reduce the effect of a correlation presence in the real signal.

## 5. Conclusions

We measured in time sequences the radioactive decay of $^{131}I$ and determined the related power-law exponent of the correlation integrals to be $\nu = 2.19 \pm 0.20$, i.e. much less than the infinite value for purely random processes. The estimated non-zero value of the Kolmogorov entropy suggests that the underlying dynamics of the physical vacuum which influences radioactive decay statistics should exhibit a deterministic chaotic mode and consequently should be fractal-like. At this point, this work should be considered as only preliminary, since several possible noise classes remain to be investigated. Future experiments utilizing simpler decay schemes and more advanced, thoroughly characterized, detector systems are required to clarify these observation. If confirmed, this would strengthen the general belief that the long-sought "theory of everything" might be sensible, since a regime of deterministic chaos in the dynamics of a system requires the existence of related deterministic equations to describe its dynamics.

## Acknowledgements

We thank Jaroslav Tobik for assistance in numerical calculation matters and Tom Girard for helpful discussions.

## References

[1] G. Goldhaber, S. Perlmutter, Phys. Rep. -Review section of Physics Letters 307 (1998) 325

[2] E. Komatsu et al., http://arxiv.org/abs/0803.0547, DOI: 10.1088/0067-0049/180/2/330

[3] J. Ambjorn, A. Gorlich, J. Jurkiewicz, R. Loll, Phys. Rev. Lett. 100 (2008) 091304

[4] D. Benedetti, Phys. Rev. Lett. 102 (2009) 111303

[5] R. Mandelbrot, Fractal Geometry of Nature, W.H. Freeman & Co., San Francisco, 1982

[6] A. S. Mikhailov: Foundations of Synergetics, Springer V., Berlin, 1990

[7] A. S. Mikhailov, A. Y. Loskutov, Foundations of Synergetics II., Springer V., Berlin, 1991

[8] E. N. Lorenz, J. Atmospheric Sci. 20 (1963) 130

[9] J. A. Yorke, Am. Math. Mon. 82 (1975) 985

[10] D. Ruelle, F. Takens, Comm. Math. Phys. 20 (1971) 167

[11] J.P. Eckmann, D. Ruelle, Rev. Mod. Phys. 57 (1985) 617

[12] E.B. Norman, S.B. Gazes, S.G. Crane, D. A. Bennett, Phys. Rev. Lett. 60 (1988) 2246

[13] E. B. Norman, B. Sur, K. T. Lesko, R. M. Larimer, D. J. DePaolo, T. L. Owens, Phys. Lett. B 357 (1995) 521

[14] R. B. Firestone, V. S. Shirley, C. M. Baglin, J. Zipkin, S. Y. F Chu, Table of Isotopes, eighth ed., Wiley-Interscience, 1996

[15] H. Kantz, Phys. Lett. A 185 (1994) 77

[16] R.H. Simonyi, A. Wolf, H.L. Swinney, Phys. Rev. Lett. 49 (1982) 245

[17] J.C. Roux, R.H. Simonyi, H.L. Swinney, Physica D, 8 (1983) 257

[18] P. Grassberger, I. Procaccia, Phys. Rev. Lett. 50 (1983) 346

[19] P. Grassberger, I. Procaccia, Physica D 9 (1983) 189

[20] P. Grassberger, I. Procaccia, Phys. Rev. A 28 (1983) 2591

[21] J. L. Kaplan, J. A.Yorke, Chaotic Behaviour of Multidimensional Difference Equations, in: Lecture Notes in Mathematics, Vol. 730, Springer V., Berlin-Heidelberg, 1979

FIG. 1 Natural logarithm of correlation integral $C_d(\varepsilon)$ vs. natural logarithm of $\varepsilon$ for the whole dataset. The values of *d* are *d* = 2 (top curve), 4, 6, ... 20 (down to bottom curve). There are no errors indicated, since calculated correlation sums are exact values.

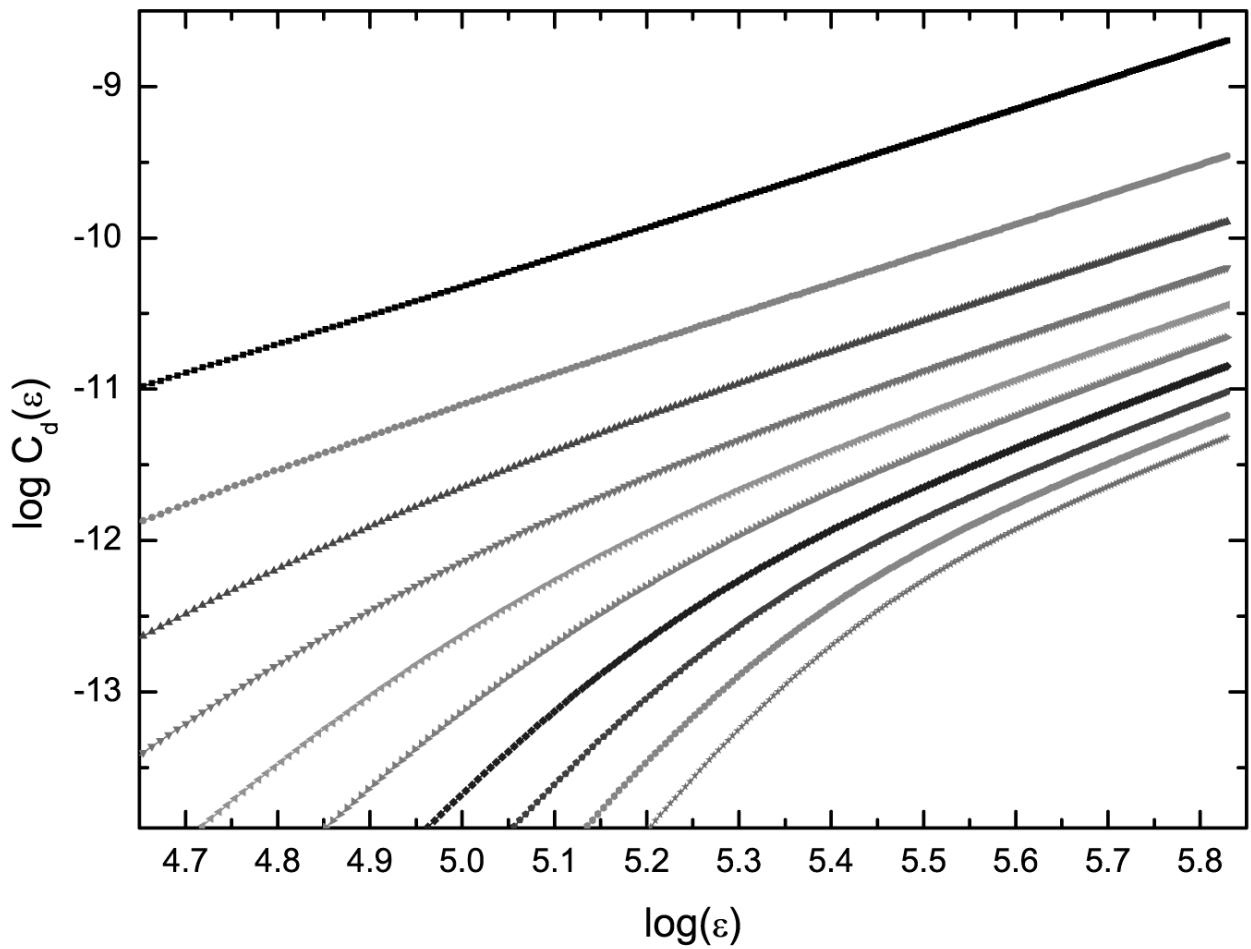

FIG. 2 The 2$^{nd}$ order Renyi entropy $K_{2,d}$ as a function of dimension d calculated at several points near the edge of the scaling region.

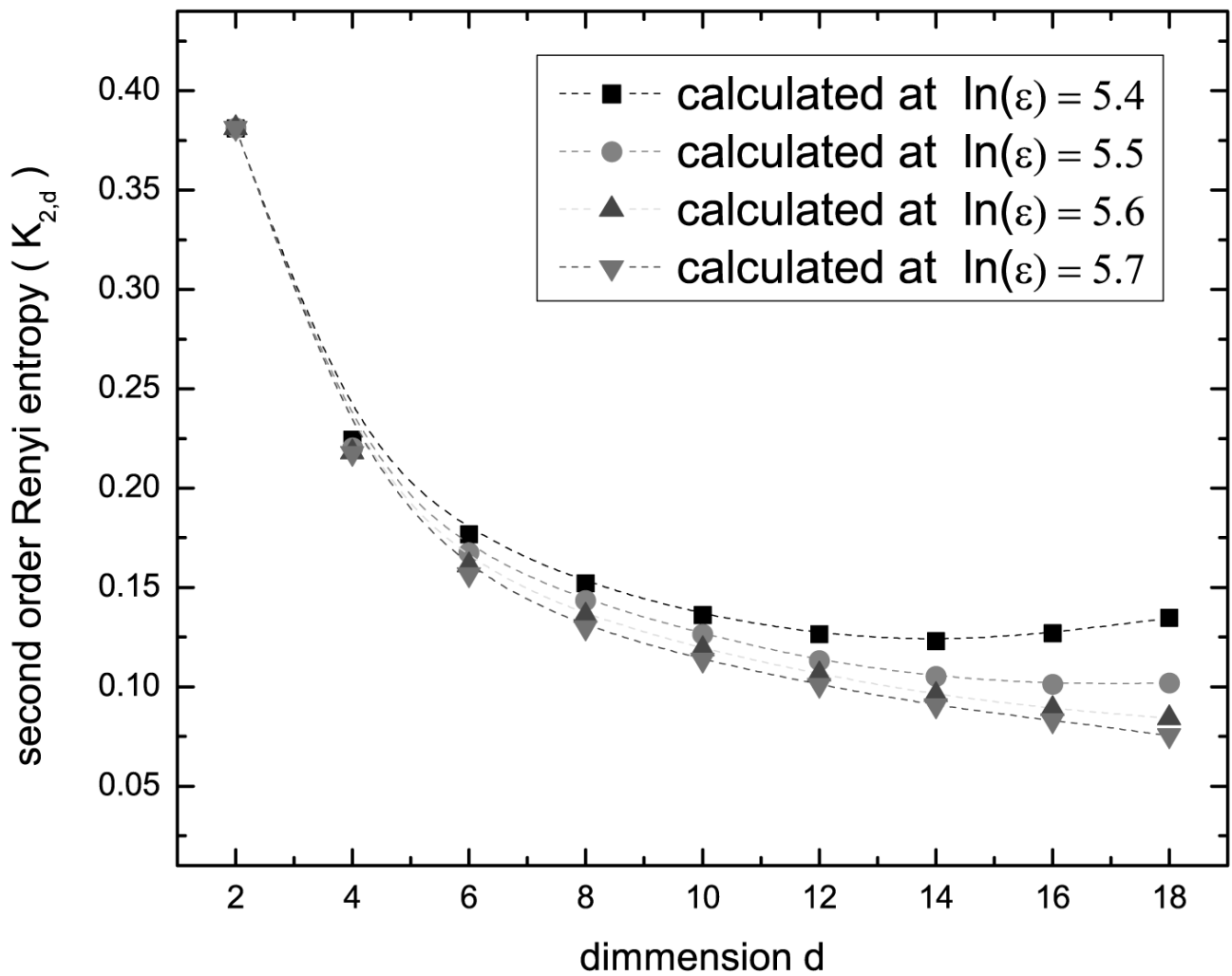